\documentclass[10pt,twoside,twocolumn,english,superscriptaddress,notitlepage,aps,prl]{revtex4-2}

\usepackage[LGR,T1]{fontenc}
\usepackage[utf8]{inputenc}
\usepackage[letterpaper]{geometry}
\usepackage{xcolor}
\usepackage{pdfcolmk}
\usepackage{babel}
\usepackage{verbatim}
\usepackage{textcomp}
\usepackage{amsbsy}
\usepackage{amstext}
\usepackage{graphicx}
\usepackage{esint}
\PassOptionsToPackage{normalem}{ulem}
\usepackage{ulem}
\usepackage[unicode=true,pdfusetitle,
 bookmarks=false,
 breaklinks=true,pdfborder={0 0 0},pdfborderstyle={},backref=false,colorlinks=true]
 {hyperref}
\hypersetup{
 colorlinks=true,citecolor=blue,linkcolor=black,urlcolor=black}
\usepackage{float} 
\usepackage{amssymb} 
\usepackage{gensymb} 

\makeatletter
\usepackage{amsfonts,anyfontsize,xcolor,graphicx}
\usepackage[calcwidth,explicit]{titlesec}
\usepackage[normalem]{ulem}

\titleformat{\section}{\bfseries\large\sffamily\scshape\filcenter}{\thesection.}{0.2em}{#1}
\titlespacing{\section}{0pt}{0.2ex}{0.2ex}
\titleformat{\paragraph}[runin]{\normalfont\normalsize\bfseries}{}{0pt}{\theparagraph}
\titlespacing*{\paragraph}{0em}{0ex}{0em}[]

\makeatletter\renewcommand\frontmatter@abstractwidth{\dimexpr\textwidth-2cm\relax}\makeatother
 
\setcitestyle{square,super}

\renewcommand\thesection{\Alph{section}}

\addto\captionsenglish{}
\AtBeginDocument{
\renewcommand{\ref}[1]{\autoref{#1}}

}

\makeatother

\begin{document}
\setlength{\abovedisplayskip}{0.2ex}\setlength{\belowdisplayskip}{0.2ex}
\setlength{\abovedisplayshortskip}{0.2ex}\setlength{\belowdisplayshortskip}{0.2ex}

\title{Chiral spin textures for next-generation memory and unconventional computing}

\author{M. S. Nicholas Tey}
\author{Xiaoye Chen}
\address{Institute of Materials Research and Engineering, Agency for Science, Technology and Research (A*STAR), Singapore}
\author{Anjan Soumyanarayanan}
\address{Institute of Materials Research and Engineering, Agency for Science, Technology and Research (A*STAR), Singapore}
\address{Physics Department, National University of Singapore, Singapore}
\author{Pin Ho}
\email{hopin@imre.a-star.edu.sg}
\address{Institute of Materials Research and Engineering, Agency for Science, Technology and Research (A*STAR), Singapore}
\begin{abstract}

\vspace{0.1cm}
\noindent
{\large\textbf{Abstract}}

\vspace{0.1cm}
\noindent 
The realization of chiral spin textures - comprising myriad distinct, nanoscale arrangements of spins with topological properties – has established pathways for engineering robust, energy-efficient and scalable elements for non-volatile nanoelectronics. Particularly, current-induced manipulation of spin textures in nanowire racetracks and tunnel junction based devices are actively investigated for applications in memory, logic and unconventional computing. In this article, we paint a background on the progress of spin textures, as well as the relevant state-of-the-art techniques used for their development. In particular, we clarify the competing energy landscape of chiral spin textures, such as skyrmions and chiral domain walls, to tune their size, density and zero-field stability. Next, we discuss the spin texture phenomenology and their response to extrinsic factors arising from geometric constraints, inter-wire interactions and thermal-electrical effects. Finally, we reveal promising chiral spintronic memory and neuromorphic devices, and discuss emerging material and device engineering opportunities.

\vspace{0.5cm}
\noindent
{\large\textbf{Keywords}}

\vspace{0.1cm}
\noindent
Memory, Computing, Spin textures, Spintronics, Chirality, Tunnel junctions, Racetrack, Neuromorphic 

\end{abstract}

\maketitle

\noindent 
\section{Introduction}
\vspace{0.3cm}

\begin{figure*}
  \includegraphics[width=\textwidth]{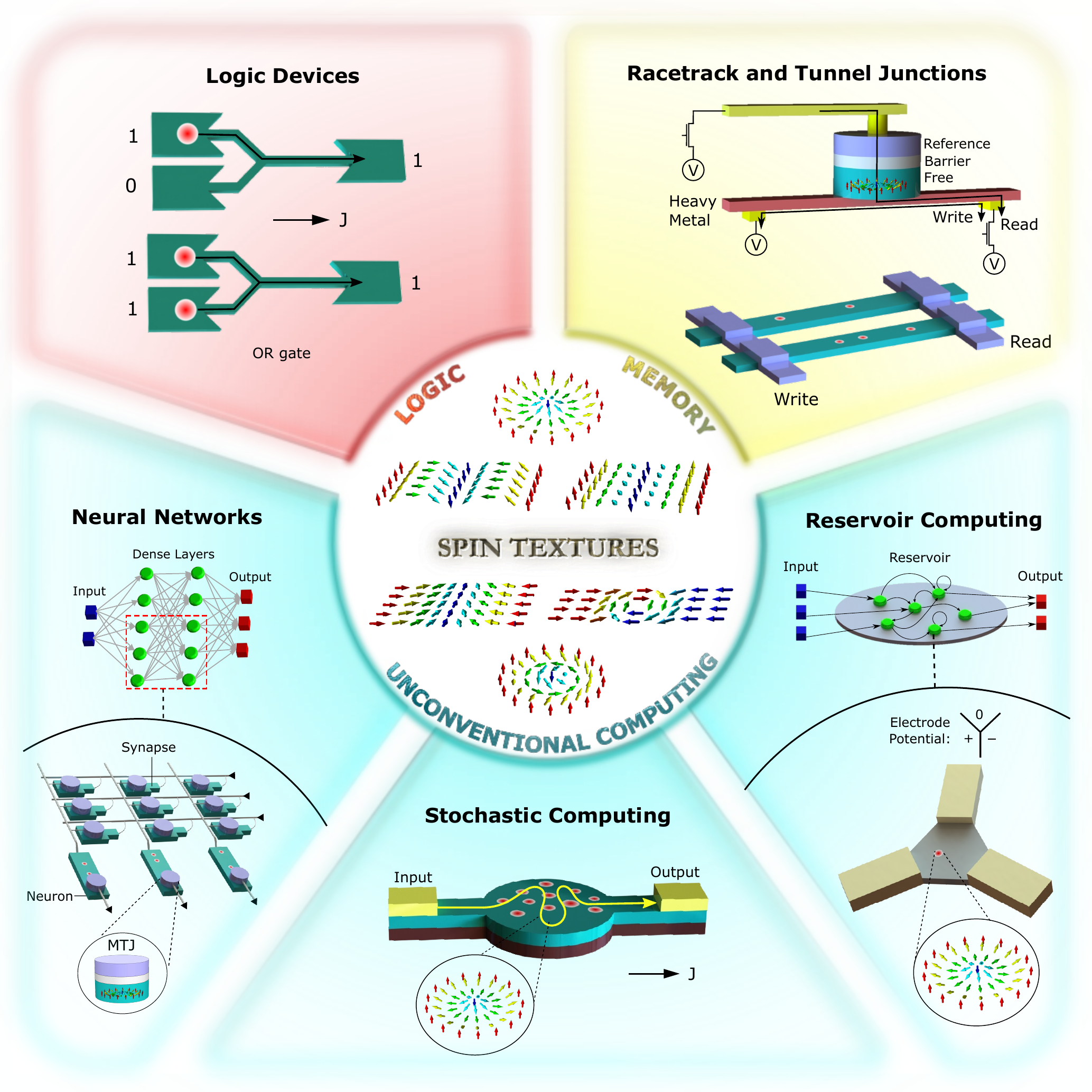}

  \vspace{-0.1cm}
  \caption{\textbf{Chiral spin memory, logic and unconventional computing.} Schematic illustration of chiral spin textures comprising left-handed Néel skyrmion, right-handed Bloch domain wall (DW), vortex DW, left-handed Bloch skyrmion, transverse DW and right-handed Néel DW (clockwise from top), and their imminent application towards non-volatile, energy-efficient and low latency technology platforms for logic (e.g. OR gates), memory (e.g. racetracks) and unconventional computing (e.g. neural networks) applications. Unconventional computing frameworks (from left to right) include neural networks (NN) comprising highly efficient brain-inspired processors (e.g. skyrmion-based synapses and neurons), stochastic computing enabling high error tolerance through probabilistic information encoding (e.g. skyrmion reshuffler devices) and reservoir computing which utilizes a fixed, non-linear system (e.g. skyrmion dynamics) to efficiently process high dimensional data.}
  
  \vspace{0.4cm}
  \label{Overview}
\end{figure*}

\paragraph{Overview}
Spintronics is a technologically and commercially relevant area of research that exploits the intrinsic magnetic spin and electronic charge of electrons for storage and memory applications. \citep{sinova2012new, nature_nanotechnology_2015} A classic representation of a spintronics device is the magnetic tunnel junction (MTJ), which consists of two ferromagnetic (FM) layers separated by a non-magnetic oxide spacer. \citep{ZHU200636, ikeda2010perpendicular} In a MTJ, binary data with variable resistance readout can be created and stored by electrically controlling the relative spin orientations of the FM layers. Till date, commercial applications of MTJs are prevalent in the read head of hard disk drives (HDD), \citep{ZHU200636, moser2002magnetic} and as memory elements in magnetic random access memory (MRAM). \citep{ZHU200636, wang2013low, nature_nanotechnology_2015} While HDD and MRAM technologies read and store binary information based on the spin orientation of static domains, new spintronic technologies have pivoted towards the use of dynamic spin textures. \citep{kumar2022domain, zheng2020paradigm} Such emerging spin texture-based technologies promise miniaturization and energy-efficient switching at smaller driving currents, setting the stage for next-generation computing applications.

\vspace{0.3cm}
\paragraph{Spin Texture Types}
Spin textures are nanometer sized structures stabilized at room temperature (RT) within crystalline and thin film materials. They comprise a variety of intriguing electron spin arrangements that separate oppositely oriented magnetic domains. First discovered as domain walls (DW) in magnetic thin films of different thicknesses, these spin textures are made up of magnetization vectors that rotate about (Bloch texture), or perpendicular to (Néel texture) the normal of the DW.\citep{yafet1988ferromagnetic, berger1992magnetic, chikazumi1997physics, hsieh2005magnetic} More recently, chiral spin textures, such as chiral DWs and skyrmions, have become a growing field of research (Fig. 1, centre). Contrary to conventional achiral Bloch or Néel DWs, chiral DWs exhibit a fixed handedness or a preferred direction for the winding of spins.\citep{chen2013novel, chen2013tailoring} Notable among these are magnetic skyrmions - topologically wound spin structures stabilized in a ferromagnetic background of opposite magnetization. The two most common skyrmion structures - Bloch and Néel-type skyrmions - are made up of spins which rotate perpendicular to or radially, respectively, from the core to the boundary. \citep{bocdanov1994properties, pollard2017observation}   

\vspace{0.3cm}
\paragraph{Spin Textures Energetics}
Within magnetic thin films, conventional spin textures are stabilized by the interplay of competing magnetic interactions - direct exchange (\textit{A}), magneto-crystalline anisotropy (\textit{K}$_\mathrm{eff}$) and magnetostatic energies - which compete to align spins in parallel, along crystallographic axes, or in the direction of the demagnetisation field, respectively. \citep{chikazumi1997physics} Meanwhile, chiral spin textures can be stabilized in heavy-metal (HM)/ferromagnet (FM) multilayer thin films with strong spin-orbit coupling and broken inversion symmetry, wherein its unique winding is attributed to an additional interfacial Dzyaloshinskii–Moriya interaction (\textit{D}, DMI) - an energetic term that favours spin canting between magnetic moments. \citep{crepieux1998dzyaloshinsky, fert2017magnetic, soumyanarayanan2016emergent} 

\vspace{0.3cm}
\paragraph{Current-Induced Torques for Spin Texture Motion}

Early predictions that a spin-polarized current can apply a torque on DW have since opened the possibility for various forms of current controlled spin texture motion,\citep{berger1978low} laying the groundwork for the burgeoning field of spin texture-based memory and computing applications. Spin textures are traditionally driven by spin transfer torque (STT),\citep{slonczewski1996current} wherein their motion is induced by the transfer of spin angular momentum, albeit at modest velocities using relatively high current densities. \citep{garello2014ultrafast, metaxas2013high} The discovery of spin orbit torque (SOT), arising from spin Hall \citep{dyakonov1971current, sinova2015spin} and Rashba-Edelstein \citep{edelstein1990spin, manchon2015new} effects, enables the generation of larger in-plane spin currents per electron charge at HM/FM interfaces. \citep{sinova2015spin, soumyanarayanan2016emergent} Consequently, superior switching speed is enabled by the exertion of such in-plane torque on the perpendicular magnetization over nanosecond timescales, \citep{baumgartner2017spatially} making SOT an increasingly attractive instrument for the electrical manipulation of chiral spin textures with ultrafast dynamics\citep{liu2012spin, miron2011fast, emori2013current, parkin2015memory}. 

\vspace{0.3cm}
\paragraph{Challenges and Article Overview}
Thus, chiral spin texture devices have several promising avenues to address critical scalability and performance requirements for next-generation memory,\citep{parkin2008magnetic, tomasello2014strategy, nakatani2016electric} logic\citep{currivan2016logic, luo2018reconfigurable} and computing applications (Fig. 1).\citep{sengupta2018neuromorphic, li2017magnetic, pinna2018skyrmion} Plausible memory devices include MTJ-like devices that incorporate spin textures for multistate readout, and racetrack memory with electrically driven spin textures representing sequential memory bits. Further, chiral hardware holds immense potential for unconventional computing frameworks such as neural networks, reservoir and stochastic computing. To this end, challenges for materials engineering and device optimization remain, and will have to be addressed to improve the performance of chiral spin texture-based memory and computing technologies. In this review, we first describe the evolution of spin textures in ferromagnetic thin films fabricated and imaged using a range of state-of-the-art microscopy techniques. Next, we provide a broad overview of their behaviour in response to extrinsic factors arising from geometric contributions, inter-wire interactions, thermal stimuli and current injections. Finally, we discuss their functionalities in racetrack memory and brain-inspired neuromorphic devices. By leveraging novel chiral material and device approaches, exciting directions for emerging chiral spintronics technologies are charted.

\vspace{0.6cm}
\section{Nanofabrication and Imaging Techniques}

\vspace{0.3cm}
\paragraph{Overview}
To elucidate the characteristics of RT chiral and achiral spin textures, a suite of experimental techniques involving film deposition, fabrication, microscopy imaging and electrical measurements are required. Here, we describe the deposition techniques used to develop a range of magnetic thin films which host RT spin textures with tunable properties, and fabrication processes for attaining nanostructures and devices in dot and wire geometries. We also introduce imaging capabilities and complementary\textit{ in situ} electrical instrumentation for identification of film and device-level spin textures along with their electrically-induced motion in devices. 

\vspace{0.3cm}
\paragraph{Deposition and Fabrication}

Typically, magnetic thin films are first deposited using ultrahigh vacuum physical vapour deposition techniques such as magnetron sputtering and ion beam deposition.\citep{ho2016domain,ho2015field,ho2017sub, tan2021intermixing, tan2021visualizing} Next, nanostructure and device patterns can be defined using electron beam lithography (EBL) and subsequently transferred using ion beam etching.\citep{ho2015field, ho2017sub, tan2021visualizing} Meanwhile, a self-assembled block copolymers (BCP) technique can be utilized to study high throughput patterning of magnetic nanowire arrays.\citep{ho2016domain, jung2010fabrication, tu2015universal} By using a carbon hard-mask patterned by a self-assembled polystyrene-block-poly(2/4-vinylpyridine) and polystyrene-blockpolydimethylsiloxane (PS-b-PDMS) diblock copolymer mask, we fabricated highly scalable, periodic nanowire arrays of sub-50 nm pitch and wire widths.

\vspace{0.3cm}
\paragraph{Spin Texture Imaging}

The magnetic force microscopy (MFM)\citep{ho2016domain,ho2015field,ho2017sub,tan2021intermixing, tan2021visualizing} and polar magneto-optic Kerr effect (MOKE) microscopy\citep{emori2013current, emori2012interfacial} are commonly utilized for the visualization and analysis of spin texture stability and dynamics. For MFM, Co-alloy coated tip scans across the sample surface at lift heights of 20-30 nm, ‘sensing’ the spin textures based on the tip-sample interactions. Tips with super sharp profile (diameter $\sim$30 nm) and ultralow magnetisation (80 emu/cm$^3$) are typically used to provide high resolution imaging with mimimal stray field perturbations. Meanwhile, polar MOKE utilizes the rotation of the polarisation of an incident beam of linearly polarised light to elucidate the magnetic structure of the sample, with a resolution limited by the wavelength of light used ($\sim$300 – 500 nm). While MFM is capable of visualising both IP and OP domain characteristics, polar MOKE enables high throughput characterization of PMA materials. Concomitant in-field imaging and electrical manipulation of spin textures can be carried out by custom designing the MFM and polar MOKE set-up to integrate magnets and electrical instruments e.g. pulse generator, amplifier and oscilloscope, and associated circuitry.\citep{ho2015field,tan2021visualizing}  

\vspace{0.6cm}
\section{Chronicle of Spin Textures}

\begin{figure*}
  \includegraphics[width=\textwidth]{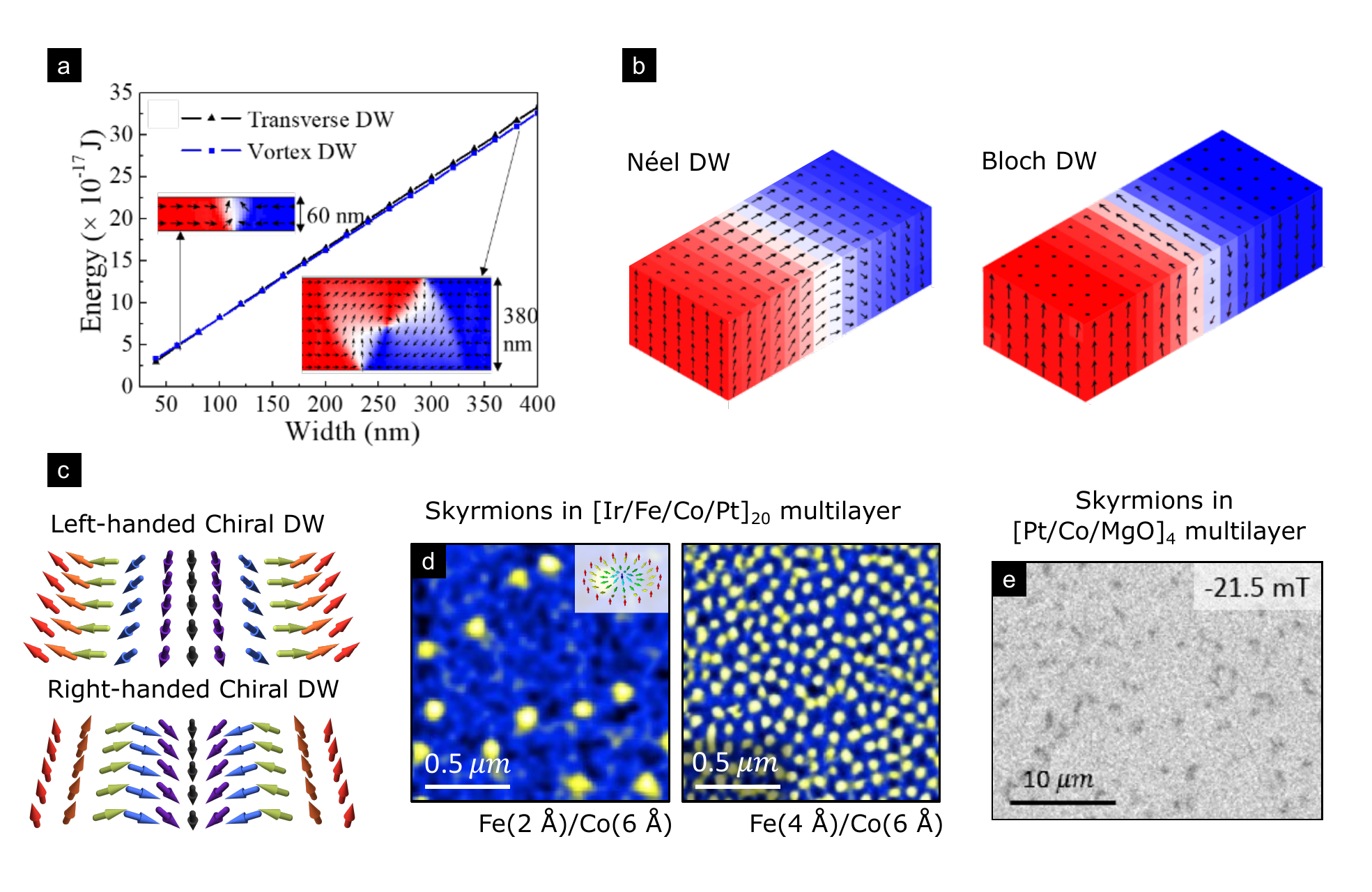}

  \vspace{-0.2cm}
  \caption{\textbf{Stabilizing spin textures.} \textbf{(a)} Achiral transverse and vortex DWs are formed in in-plane anisotropy (IPA) material systems such as A1-FePt. Vortex DWs are stabilized above a critical wire width. (Reproduced from P. Ho \textit{et al.}\citep{ho2015field} Copyright 2015, IEEE.) \textbf{(b)} Achiral Néel and Bloch DWs stabilized in perpendicular magnetic anisotropy (PMA) material platforms such as multilayered [Co/Pd]$_{15}$ and \textit{L}1$_\mathrm0$-FePt. (Reproduced from P. Ho \textit{et al.}\citep{ho2016domain} Copyright 2016, The Royal Society of Chemistry.) \textbf{(c)} Chiral DWs with left-handed and right-handed chirality are stabilised depending on the sign of the Dzyaloshinskii–Moriya interaction (DMI) \textbf{(d)} Skyrmions stabilized in a [Ir/Fe/Co/Pt]$_{20}$ multilayer, wherein DMI from the Fe/Ir and Co/Pt interfaces act in concert to enhance the effective DMI. At $H \approx -0.8$ $H_\mathrm{S}$, magnetic force microscopy (MFM) images (scale bar: 0.5~μm) show isolated (left) and lattice (right) skyrmion configurations favoured by tuning the Fe and Co thicknesses. Inset shows the schematic of a Néel skyrmion. \textbf{(e)} Skyrmions stabilized in [Pt/Co/MgO]$_{4}$ multilayers at $H \approx -0.25$ $H_\mathrm{S}$, imaged using polar magneto-optic Kerr effect microscopy (scale bar: 10~μm).}
  \vspace{0.4cm}
  \label{Section2}
\end{figure*}

\vspace{0.3cm}
\paragraph{Achiral Spin Textures}

Conventional, or achiral spin textures are found in both in-plane anisotropy (IPA) (e.g. permalloy, A1-FePt, etc.) and perpendicular magnetic anisotropy (PMA) magnetic materials (e.g. Co/Pd, \textit{L}1$_\mathrm0$-FePt, etc.) wherein interfacial DMI contributions are absent. In IPA thin films, vortex and transverse DWs are favoured in wide and narrow wires, respectively (\ref{Section2}a). \citep{ho2015field,zhang2015effects,zhang2016360, mcmichael1997head} Vortex spin structures adopt a two-fold rotational symmetry about their centre, while transverse DWs are characterized by a reflection symmetry about the normal of the DW. \citep{mcmichael1997head} STT-driven motion of 180° transverse DWs in narrow nanowires have limited practical use as the maximum achievable velocity is constrained by the onset of Walker Breakdown, a critical current density beyond which the DW structure periodically switches between the transverse and vortex configurations.\citep{beach2008current, zhang2015effects, zhang2016360} In contrast, a metastable 360° transverse DW, formed through the attraction of two 180° DWs of opposite core magnetisation, offers topological protection and circumvents Walker Breakdown.\citep{zhang2016360, mascaro2010ac} Meanwhile, achiral Néel (\ref{Section2}b, left) or Bloch (\ref{Section2}b, right) DWs can be stabilized in PMA thin films.\citep{ho2016domain} In [Co/Pd]$_{15}$ multilayered nanowires, we observed that the low spin torque efficiency, coupled with a limited threshold current density to minimise joule heating effects, led to performance bottlenecks for deterministic and fast achiral DW dynamics.\citep{ho2016oersted}

\vspace{0.3cm}
\paragraph{Chiral Spin Textures}

In this light, RT chiral spin textures, consisting of a large range of topological spin textures with fixed handedness (\ref{Section2}c-e), have gained traction for practical device implementation due to their energy-efficient SOT-driven dynamics. \citep{emori2013current, soumyanarayanan2016emergent, miron2011fast, emori2012interfacial} These textures have been stabilised in various HM/FM and HM/FM/oxide multilayers including the pinning-free Pt/CoB/Ir, Ir/CoB/Pt and Pt/CoB/MgO amorphous material platforms.\citep{tan2021intermixing} Here, we reported that PMA can be enhanced through controlled intermixing within the metallic Pt/CoB/Ir and Ir/CoB/Pt stacks, and  thermal annealing across the three amorphous stacks. This provides a means of \textit{K}$_\mathrm{eff}$ tuning for a range of chiral spintronics-based technological applications.

\vspace{0.3cm}
\paragraph{Chiral Spin Textures: Skyrmions}
Given that skyrmions are notable manifestations of these RT chiral domain walls, \citep{soumyanarayanan2016emergent} we developed a tunable RT skyrmion platform based on the multilayer stacks of [Ir/Fe/Co/Pt]$_{20}$.\citep{soumyanarayanan_tunable_2017} By harnessing the large and opposite signs of DMI generated from Fe/Ir and Co/Pt interfaces, we demonstrated that the additive enhancement of DMI can stabilize skyrmions with diameters as small as 30 nm. By further varying the Fe and Co thicknesses, the competing magnetic parameters \textit{D}, \textit{K}$_\mathrm{eff}$ and \textit{A} can be modulated to tune the skyrmion size (2x) and density (10x). Crucially, by tuning these magnetic parameters, we illustrated a smooth crossover from a sparsely populated skyrmion configuration (\ref{Section2}d, left) to a short-range hexagonally ordered skyrmion lattice (\ref{Section2}d, right). This enables the creation of designer skyrmion configurations which can be harnessed for independent lines of device applications.

\vspace{0.6cm}
\section{Geometric Effects}

\begin{figure*}
  \includegraphics[width=\textwidth]{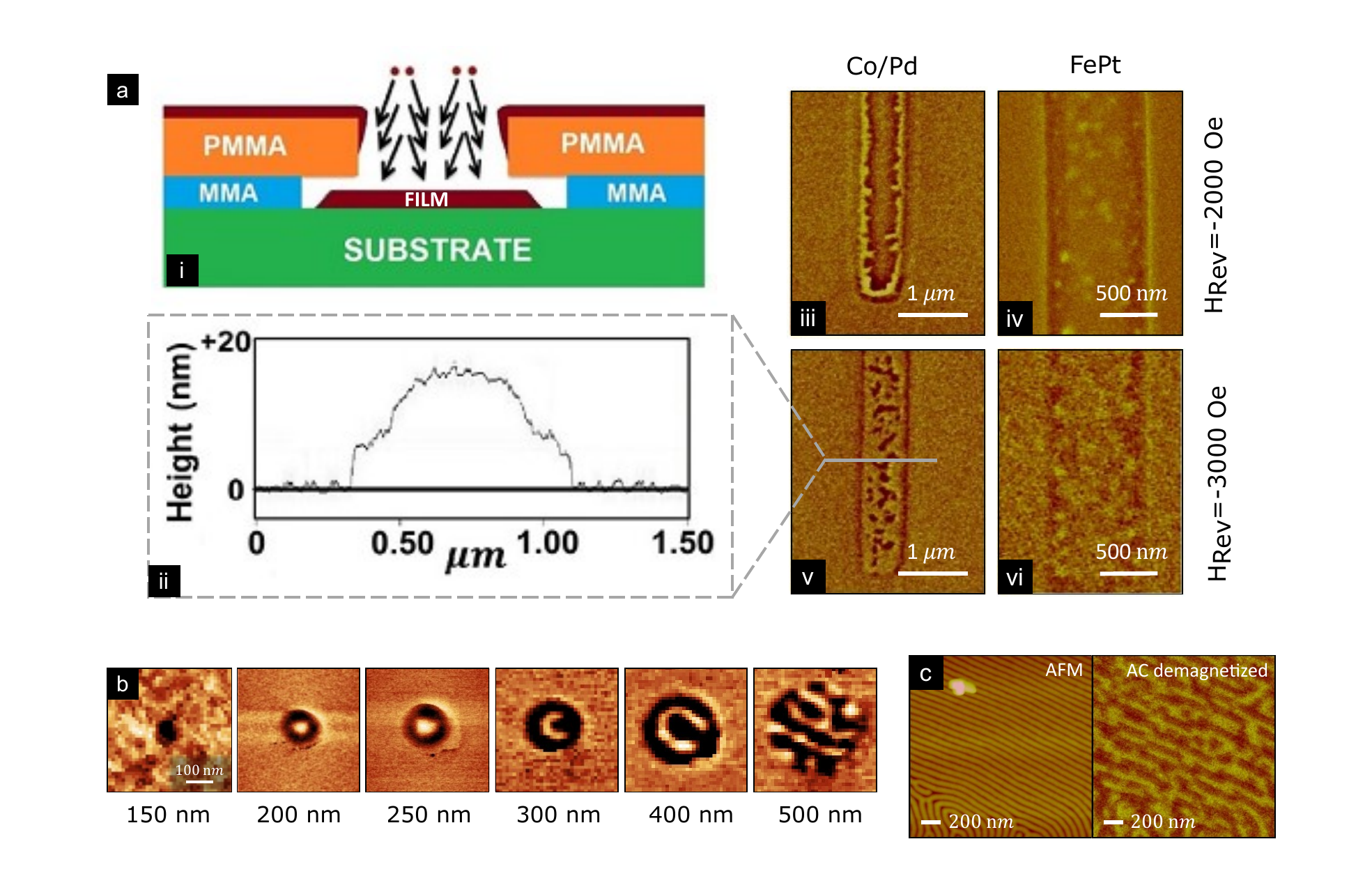}
  \vspace{-0.6cm}

  \caption{\textbf{Geometric influences on spin textures.} \textbf{(a-i)} Wire profile with an edge taper is achieved via lift-off using a double-layer resist with an undercut profile. (ii) Cross-sectional profile of a representative, tapered nanowire measured using atomic force microscopy. (iii-vi) MFM imaging of [Co/Pd]$_{20}$ (scale bar: 1~μm) and \textit{L}\(1_\mathrm{0}\)-FePt (scale bar: 500 nm) tapered nanowires at reversal fields of (iii, iv) $H_{Rev}= -2000$ Oe and (v-vi) $-3000$ Oe, after saturation at +10 kOe, show distinct edge-modulated reversal behaviour. Initial nucleation is favoured at the edge and centre of [Co/Pd]$_{20}$ and \textit{L}\(1_\mathrm{0}\)-FePt wires, respectively. (Reproduced from J. Zhang \textit{et al.} \citep{zhang2015edge} Copyright 2015, AIP Publishing LLC.) \textbf{(b)} MFM images (scale bar: 100 nm) of [Ir/Fe(3 Å)/Co(6 Å)/Pt]$_{20}$ nanodots show confined zero field skyrmions at intermediate dot sizes, with uniformly magnetized and labyrinthine stripe phases for smaller and larger dot sizes respectively. (Reproduced from P. Ho \textit{et al.}\citep{ho2017sub} Copyright 2019, American Physical Society.) \textbf{(c)} AFM image (scale bar: 200 nm) of the [Co/Pd]$_{15}$ nanowire array. Corresponding MFM image (scale bar: 200 nm) illustrates the preferential anti-correlated magnetization configuration arising from inter-wire dipolar stray fields upon AC demagnetisation. (Reproduced from P. Ho \textit{et al.}\citep{ho2016domain} Copyright 2016, The Royal Society of Chemistry.) }
  \vspace{0.4cm}
  \label{Section3}
\end{figure*}
\vspace{0.3cm}
\paragraph{Wire Taper and Edge}

Geometric factors such as the tapering of wire profiles and size variations are critical to the stability and dynamics of spin textures, as it affects the homogeneity of magnetic parameters,\citep{zhang2015effects, maranville2006characterization, zhang2015edge, lai2017improved, yoo2017current} introduces pinning \citep{zhang2016360, zhang2015edge, GR_Jiang, reichhardt2021statics} and confinement effects. \citep{rohart2013skyrmion, ho2017sub} Edge taper, characterised by a trapezoidal wire cross-section, is a common feature of patterned wires due to the shadowing effect that arises from the etching process (\ref{Section3}a). \citep{drotar2000surface,zhang2015edge,zhang2015effects} In IPA nanowires hosting transverse DWs, we reported that tapering can modify demagnetisation fields at the edge and delay the onset of Walker Breakdown, thereby increasing the maximum DW velocity by a factor of $\approx2$. \citep{zhang2015effects} In PMA nanowires, the \textit{K}$_\mathrm{eff}$ of the tapered wire can be tuned to allow spin texture nucleation at desirable switching fields (\ref{Section3}a). \citep{zhang2015edge} This effect is material dependent - preferential spin texture nucleation is observed at the tapered edges of [Co/Pd]$_{15}$  (\ref{Section3}aiii, v) due to their degraded interfacial-induced PMA, while initial nucleation is favored at the center of the \textit{L}\(1_\mathrm{0}\)-FePt wire (\ref{Section3}aiv, vi) due to higher PMA at the thinner edge. Additionally, we observed that edge-induced pinning effects impeding DW translation became significantly detrimental for narrower wires (width: $\le$ 50 nm). \citep{zhang2016360} 

\vspace{0.3cm}
\paragraph{Geometric Confinement}
Geometric confinement of chiral multilayer thin films also dictates the ambient stability and evolution of skyrmions. We observed the zero field (ZF) stability of [Ir/Fe/Co/Pt]$_{20}$ skyrmions over a critical range of geometric sizes (200 – 300 nm)  (\ref{Section3}b). \citep{ho2017sub} While the chiral stripe phase dominates at large dot sizes, skyrmions are favoured for small dot sizes due to a reduction in magnetostatic energy. The ZF skyrmion size can be modulated systematically by a factor of 4, down to 50 nm, by varying geometric size and magnetic interactions. Such geometric tuning, enabling their small size and ZF stability, allows for the design of compact devices compatible with existing electrical architectures. 

\begin{figure*}
  \includegraphics[width=\textwidth]{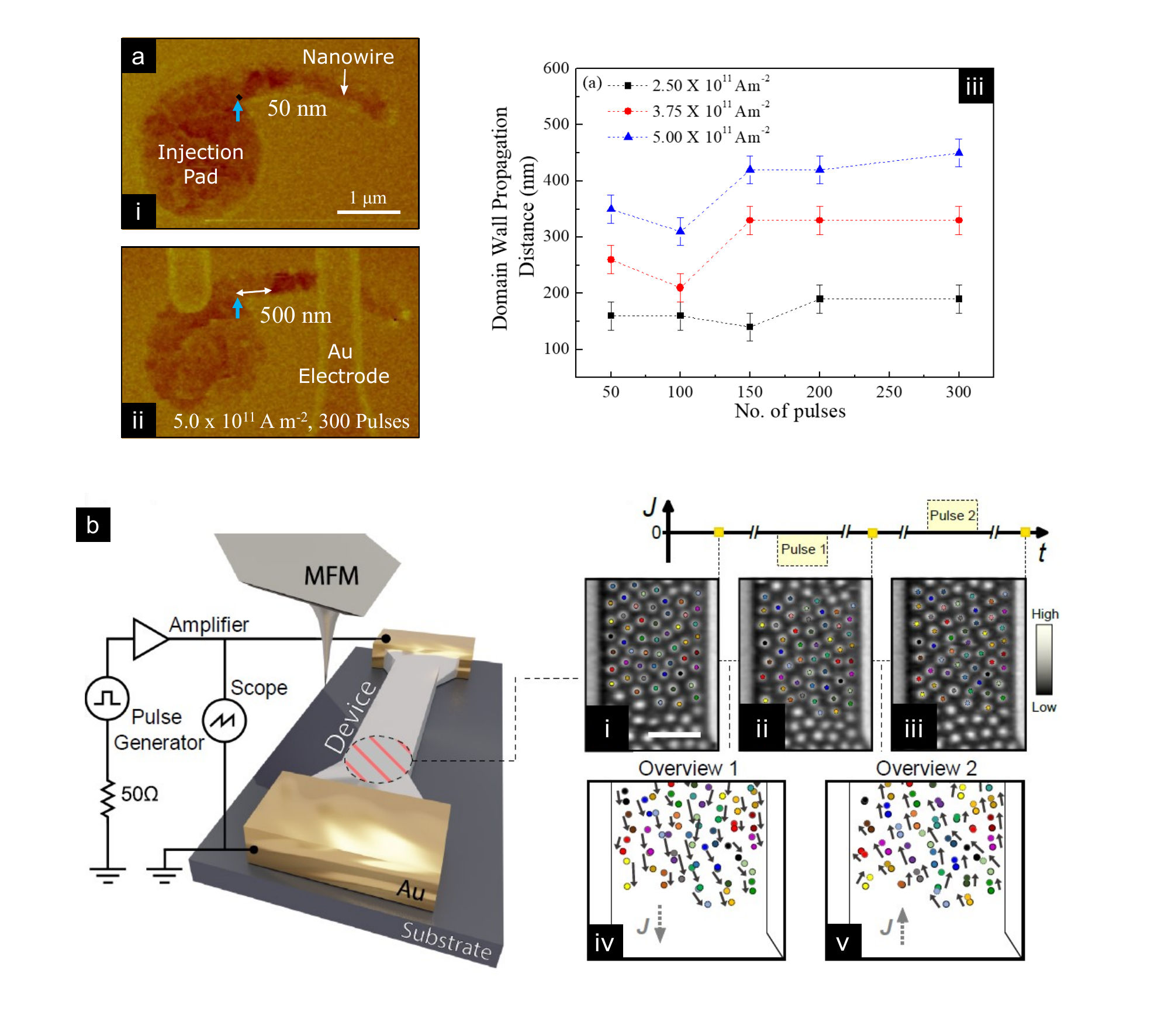}
  \vspace{-0.9cm}
    
 \caption{\textbf{Electrical Manipulation of spin textures.} \textbf{(a)} MFM images of arc-shaped A1-FePt nanowire (scale bar: 1~μm) indicating (i) initial and (ii) final vortex DW positions after injecting 300 pulses of current density $5\times10^{11}$ $\mathrm{Am^{-2}}$ with pulse width of 10~μs. (iii) DW propagation distance as a function of number of applied pulses across current densities of $(2.5 - 5)\times10^{11}$ $\mathrm{Am^{-2}}$. (Reproduced from P. Ho \textit{et al.}\citep{ho2015field} Copyright 2015, IEEE.) \textbf{(b)} Schematic of the MFM and electrical pulsing setup with varying \textit{in situ} out-of-plane magnetic fields for examining the ensemble dynamics of individual skyrmions in a [Pt/Co/MgO]$_{15}$ multilayered wire device. Pulses of opposite polarity ranging from $\pm(1 - 5.8)\times10^{11}$ $\mathrm{Am^{-2}}$ were applied sequentially for 20 ns intervals, and MFM images of the wire (i-iii, scale bar: 1~μm) were acquired before and after each pulse. Skyrmions are colour-tagged and individual skyrmion motion are tracked. (iv-v) Arrows representing their motion due to current pulses of opposite polarity. (Reproduced from A. K. C. Tan \textit{et al.}\citep{tan2021visualizing} Copyright 2021, Springer Nature.)} 
  \vspace{0.4cm}
  \label{Section4}
\end{figure*}

\vspace{0.3cm}
\paragraph{Inter-wire Interactions}

Inter-wire interactions, namely magnetostatic interactions between closely packed nanowire arrays, is critical to device scalability.\citep{wu2020impact, yoon2017modeling} In BCP patterned [Co/Pt]$_{15}$ and \textit{L}1$_\mathrm{0}$-FePt nanowire arrays with $\sim$ 60 nm pitch and 30 - 40 nm width, domain stability is observed to be dependent on the competing effects of PMA and magnetostatic interactions. \citep{ho2016domain} While dipolar stray fields are considered a long range effect, it is negated at sub-100 nm spacings for \textit{L}\(1_\mathrm{0}\)-FePt nanowires due to their relatively high coercivity. For less coercive [Co/Pd]$_{15}$ nanowires, however, dipolar stray fields favoured the propagation and stabilisation of anti-correlated domains between adjacent wires (\ref{Section3}c). Alternating magnetization configuration is observed in $\sim$75\% of the wires, while parallel-aligned magnetization in the remaining wires is attributed to frustration, producing degenerate ground states susceptible to small perturbations. The non-trivial effects of geometric influences such as edge taper, nanostructure spacings and confinement underscore the importance of geometric considerations for device applications.

\vspace{0.6cm}
\section{Electrical Effects}
\vspace{0.3cm}
\paragraph{Electrical Manipulation of Achiral Textures}
When subjected to lateral currents, spin torques can induce fast spin texture motion of close to 1000 m/s.\citep{parkin2015memory} Early works examined the current driven motion of DWs in IPA nanowire devices made of permalloy, Co and A1-FePt. (\ref{Section4}a). \citep{ho2015field, zhang2016360} While STT-induced DW velocity can be increased substantially (>3$\times$) by applying an IP field ($\sim$100 Oe), \citep{ho2015field} driving DWs with the simultaneous use of an external field and current is incompatible with prevailing technologies. 

\vspace{0.3cm}
\paragraph{Electrical Manipulation of Chiral Textures}
More recently, efforts are directed at demonstrating the SOT-induced dynamics of topologically protected spin textures. We analyzed the current-driven ensemble dynamics of 80 – 200 nm sized skyrmions in [Pt/Co/MgO]$_{15}$ multilayered wires by examining over 20,000 instances of individual skyrmion motion across currents and fields (\ref{Section4}b). \citep{tan2021visualizing} Notably, the skyrmion velocity increases exponentially with current densities, reaching as high as 24 m/s. There is a marked transition in dynamic regimes with increasing current – from stochastic to deterministic creep, before concomitant motion of skyrmions in the plastic flow regime. In addition to the expected linear motion in the plastic flow regime, the current also induced a transverse deflection of $\sim$22°. Known as the skyrmion Hall effect (SkHE), the deflection arises from the Magnus force acting on the skyrmionic topological charge. While SkHE enables defect avoidance, the non-linear motion can also result in skyrmion-edge interactions. At the wire edge, we observed that skyrmions exhibit weakly inelastic interactions and are not annihilated or pinned. Importantly, small skyrmions move with smaller deflection angles, while their velocities are comparable to larger skyrmions. Particle model simulations suggest that such SkHE size dependence arises from pinning-driven effects. \citep{reichhardt2021statics, tan2021visualizing} 

\vspace{0.3cm}
\paragraph{Joule Heating}
Joule heating is an undesirable by-product of pulsed electrical currents which, in some cases, can lead to the destruction of ferromagnetic order and is often deleterious to device reliability. \citep{yamaguchi2005effect, moretti2016influence} Early work showed that the temperature rise across a nanowire increases with the pulse width and magnitude of the injected current, \citep{ho2016oersted, ho2015field} and permanent damage arising from insufficient heat dissipation can manifest as an overall increase in the resistance of the wire sample. \citep{ho2016oersted} In the presence of considerable Joule heating, non-directional DW motion and random nucleation of domains are also observed.\citep{ho2016oersted} Meanwhile, Joule heating effects can be contained by using shorter pulse widths and lower current magnitudes, or by increasing the time interval between pulses. We reported a modest rise in temperature of $\sim$3.3 K in A1-FePt nanowires by suitably controlling the current density (<$\sim$\(5\times 10^{11}~\mathrm{Am}^{-2}\) and pulse width (<10~μs).\citep{ho2015field} Ultimately, to build a robust and energy-efficient device framework compatible with prevailing spintronics architectures, understanding the subtleties and limitations specific to current driven manipulation of spin textures is paramount.

\vspace{0.6cm}
\section{Chiral Spintronic Technologies}

\begin{figure*}
  \includegraphics[width=0.95\textwidth]{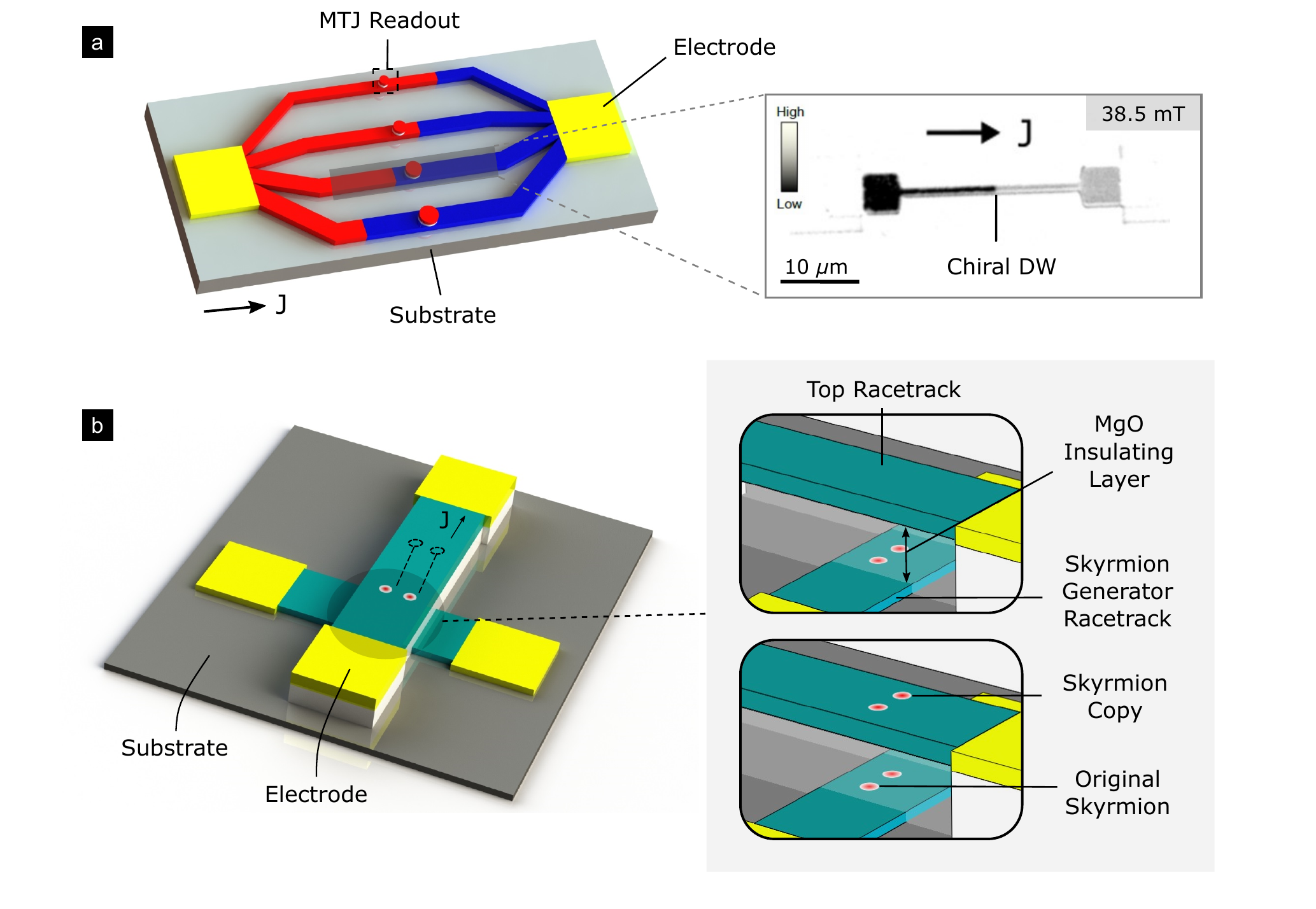}
  \vspace{-0.1cm}
  
  \caption{\textbf{Neuromorphic computing concepts using spin textures.} \textbf{(a)} Illustration of a chiral DW-based synaptic device built upon the concept of a multi-state variable resistor. Chiral DWs are driven through the array of racetracks using an injected SOT current and a magnetoresistance readout is detected by the magnetic tunnel junction (MTJ). The varying racetrack widths modulate the current densities and extent of DW motion in each racetrack. Inset shows the current driven motion of DWs in a representative racetrack imaged by MOKE (scalebar: 10~μm). A multi-state variable resistor is obtained by measuring the collective resistance across all MTJ readouts. \textbf{(b)} Schematic of a skyrmion-based synaptic device, comprising two perpendicularly stacked racetracks separated by an insulating MgO layer. The synaptic weight is determined by the number of skyrmions at the racetrack intersection. The inset illustrates the process before (top) and after (bottom) skyrmion copies are created as a result of stray fields from skyrmions in the generator racetrack. Synaptic multiplication is performed since the output signal (i.e. number of skyrmions in the top racetrack) is derived from the product of the weight and the input signal, the latter being the number of current pulses through the top racetrack. }
  \vspace{0.4cm}
  \label{Section5}
\end{figure*}
\vspace{0.3cm}
\paragraph{Racetrack Memory/Logic}

Devices with racetrack architectures hold promise as a fast, compact, energy-efficient and non-volatile memory solution\citep{wang2018ultra, fert2017magnetic, soumyanarayanan2016emergent}. By design, the bit length is defined by regularly spaced pinning sites fabricated along a magnetic nanowire, and spin polarised current pulses are used to drive a sequence of magnetic bits or DW across a write and read head.\citep{parkin2008magnetic, parkin2015memory} Our work forms the cornerstone for functional chiral racetrack device implementation by addressing various fronts pertinent to writing, switching and reading operations. First, we demonstrated deterministic writing of DWs using a current-induced Oersted field generated from a Au wire fabricated perpendicular to the racetrack. \citep{ho2016oersted, ho2015field} The magnetization orientation of the nucleated domain can be controlled by the polarity of current pulse through the Au wire. Second, we demonstrated the SOT-induced dynamics of a dense array of sub-100 nm skyrmions in wire devices, paving the way for materials and device directions involving high-throughput racetracks.\citep{tan2021visualizing} Notably, their modest SkHE, comparable to that in ferrimagnets, suggest that materials design efforts may continue in the direction of ferromagnets. Amorphous multilayer thin films hosting chiral spin textures have also been developed for their enhanced dynamics and compatibility with MTJ integration for detection. \citep{tan2021intermixing} Coupled with ongoing efforts to stabilize RT skyrmions at ZF, our work is poised to support the development of next-generation racetrack devices hosting novel spin textures.

\vspace{0.3cm}
\paragraph{Neural Computing Introduction}
To realise edge intelligence, proposals for in-memory computing to completely overhaul the current von-Neumann architecture has gained immense interest. Brain-inspired neuromorphic computing using hardware-based neural networks (NN), comprising billions of neurons interconnected by an even larger number of synapses, aims to increase the efficiency of computationally intensive tasks like machine learning and pattern recognition. \citep{lawrence2021matrix, le2018mixed} Chiral spintronics hardware platforms are well suited to execute such tasks, as they are non-volatile, energy-efficient and exhibit a mixture of digital and analogue characteristics. Specifically, the solitonic, weakly interacting nature of skyrmions allow them to represent the indistinguishable, binary spike signals in a spiking NN, while their spatiotemporal current-driven motion can be used to intuitively track spike signal transmissions characteristic of information stored in the time domain (\ref{Overview}). \citep{intro_Song, intro_Diehl}

\vspace{0.3cm}
\paragraph{Neural Computing Concepts}
As a key component of neuromorphic devices, synapses require tunable and non-volatile analog weights, with a function enabling multiplication between the weights and an input signal. One such synaptic device concept involves the use of a current output signal, which is itself the product of a variable resistor weight and a fixed input voltage (\ref{Section5}a). In this approach, a write current will deterministically move a chiral DW, providing fine control of the DW position and the resultant magnetoresistance readout. We posit that a variable resistance can be achieved by simultaneously driving chiral DWs in an array of racetracks with variable widths. Another concept is built upon the creation of skyrmion copies in an architecture consisting of two skyrmion-hosting racetracks stacked perpendicularly in a cross-bar manner (\ref{Section5}b).\citep{privateCom} Skyrmion copies in the top racetrack are replicated from the stray fields of skyrmions in the skyrmion generator racetrack, and subsequently sheared away upon injecting a sufficiently large current in the top racetrack. Synaptic multiplication is realised via an output signal i.e. number of sheared skyrmion copies, which is represented by the product of the synaptic weight and input pulses through the top racetrack. Synaptic weights can be modified by varying the number of skyrmions in the intersection region of the skyrmion generator. 

\vspace{0.6cm}
\section{Conclusion}

\vspace{0.3cm}
\paragraph{Summary}

In summary, we have presented a comprehensive picture on the advances of spin textures from IPA (transverse, vortex) and PMA (Néel and Bloch) DWs, to chiral spin textures (chiral DW, skyrmions) with topological spin arrangements. Cognizant of the limitations of achiral spin textures such as large STT current densities and susceptibility to Walker Breakdown, research focus is now geared towards harnessing chiral spin textures that can be manipulated using fast, energy-efficient SOT. By fine-tuning the magnetic parameters via material stack design, nanostructure engineering and thermal annealing processes, formation of room-temperature chiral spin textures with tunable physical properties such as size, density and stability can be realised. Additionally, we established a robust framework to harness the dynamics of high-throughput chiral spin textures, in particular skyrmions, for next-generation memory and computing technologies. We underscored the suitability of such chiral spin textures for artificial synapses and neurons in NNs, and also discussed practical device-level considerations, including the effects of joule heating, inter-wire stray field interactions and geometric confinement, which can serve as design guides for future chiral spintronics devices. Particularly, these insights serve as an important stepping stone for realising disruptive edge intelligence technologies, including racetrack memory, logic devices and unconventional computing elements.

\vspace{0.3cm}
\paragraph{Outlook - Materials}

In tandem with the aforementioned developments, notable strides have been made on related materials front. First, the synthetic antiferromagnetic multilayer has been hailed as a practical material platform enabling significantly enhanced DW velocity ($\sim$1000 ms$^{-1}$) \citep{parkin2015memory, yang2015domain} and linear skyrmion motion (zero SkHE) by the mutual compensation of topological charges. \citep{zhang2016magnetic, li2020artificial, legrand2020room} Second, ferrimagnetic material systems \citep{hirata2019vanishing, woo2018current} can offer alternative means of suppressing SkHE, albeit requiring low temperature operations ($\sim$275 K) for optimal velocity and zero SkHE. \citep{hirata2019vanishing} Third, an emergent class of chiral antiferromagnets, with topology arising from their intrinsic non-collinear spin order, are competitive to prevailing chiral ferromagnetic counterparts due to their zero parasitic stray fields and fast switching dynamics. \citep{takeuchi2021chiral, li2019chiral, gomonay2018antiferromagnetic,nakatsuji2015large} Finally, topological multiferroic textures, integrating both ferroelectric and ferromagnetic functionalities, enable energy-efficient voltage-induced switching in chiral spintronic devices and is poised to be more relevant as the drive towards miniaturisation intensifies.\citep{kim2018configurable} These diverse material platforms chart an exciting roadmap for the promising future of chiral spintronics research and applications. 

\vspace{0.3 cm}
\paragraph{Outlook - Devices}
On the device front, directions focusing on 3D chiral architectures, embodying vertical stacking of chiral DWs or skyrmion based-MTJs \citep{kim2011stacked} and racetracks \citep{parkin2008magnetic}, may pave the way towards maximizing areal efficiency of chiral spintronic technologies. Future studies on spin-based stochastic computing may also find that, in the presence of thermal stimuli, the inherently stochastic nature of DWs and skyrmions can be used to implement an error-resistant computing framework, \citep{zhang2020stochastic} or enhance the bio-realistic nature of hardware accelerated neural networks \citep{lone2021voltage}. Reservoir computing initiatives are also poised to exploit the non-linear dynamics of DWs \citep{ababei2021neuromorphic} and skyrmions,\citep{pinna2020reservoir} potentially improving the processing performance of high dimensional data. Additionally, topological quantum computing is expected to similarly outperform conventional computational models and traditional quantum computers for certain types of problems.\citep{gyongyosi2019survey} In this vein, skyrmions play an important role as they host Majorana bound states – superconducting vortex pairs in 2-dimensional structures – that enable robust, scalable and electrically addressable qubit operations.\citep{nothhelfer2022steering, yang2016majorana, ma2020braiding} The striking advantages of skyrmion qubits and multiqubits pave promising directions for quantum computing. Even at this nascent stage, seminal advances in chiral spintronic technologies point towards an imminent revolution of traditional computing frameworks.

\vspace{0.6cm}
\textbf{Acknowledgement} 

\vspace{0.2cm}
This work was supported by the SpOT-LITE programme (Grant No. A18A6b0057), funded by Singapore’s RIE2020 initiatives, and by the Career Development Fund (Grant No. C210812017) funded by A*STAR, Singapore.

\phantomsection\addcontentsline{toc}{section}{\refname}

\bibliography{References}
\bibliographystyle{achemso}

\noindent \begin{center}
{\small{}\rule[0.5ex]{0.6\columnwidth}{0.5pt}}{\small\par}
\par\end{center}

\newpage
\onecolumngrid

\makeatletter
\makeatother
\setcounter{equation}{0}
\setcounter{figure}{0}

\twocolumngrid

\end{document}